\documentclass[12pt]{iopart}
\usepackage{iopams, graphics, epsfig, soul}
\def\be{\begin{equation}}
\def\ee{\end{equation}}
\def\bea{\begin{eqnarray}}
\def\eea{\end{eqnarray}}
\def\l{\left}
\def\r{\right}

\def\det{\sqrt{-g}}
\def\R{\mathcal{R}}
\def\4{\;\;\;\;}
\def\2{\;\;}
\def\a{\alpha_}
\def\b{\beta_}
\def\ta{\tilde{\alpha}_}
\newcommand{\s}{\stackrel}
\newcommand{\nn}{\nonumber}

\begin{document}
\title{The role of nonmetricity in metric-affine theories of gravity}

\author{{Vincenzo Vitagliano}}

\address{CENTRA, Departamento de F\'isica, Instituto Superior T\'ecnico,
Universidade T\'ecnica de Lisboa - UTL, Av. Rovisco Pais 1, 1049 Lisboa, Portugal}

\ead{\mailto{vincenzo.vitagliano@ist.utl.pt}}

\begin{abstract}
The intriguing choice to treat alternative theories of gravity by means of the Palatini approach, namely elevating the affine connection to the role of
independent variable, contains the seed of some interesting (usually under-explored) generalizations of General Relativity, the metric-affine theories 
of gravity. The peculiar aspect of these theories is to provide a natural way for matter fields to be coupled to the independent 
connection through the covariant derivative built from the connection itself. Adopting a procedure borrowed from
the effective field theory prescriptions, we study the dynamics of metric-affine theories of increasing order, that in the complete version include
invariants built from curvature, nonmetricity and torsion. We show that even including terms obtained from nonmetricity and torsion to the second order 
density Lagrangian, the connection lacks dynamics and acts as an auxiliary field that can be algebraically eliminated, resulting in some extra 
interactions between metric and matter fields.\\
{\centerline{\hspace{-2.5cm}\em Dedicated to the memory of Francesco Caracciolo}}
\end{abstract}

%%%%%%%%%%%%%%%%%%%%%%%%%%%%%%%%%%%%%%%%%%%%%%%%%%%%%%%%%%%%%%%%%%%%%%%%%%%%%%%%%%%%%%%%%%%%%%%%%%%%%%%%%%%%%%%%%%%%%
\section{Introduction}

Despite the name ``Palatini approach'', Einstein himself recognized the possibility to formally describe the theory of General Relativity 
assuming the independence of the affine connection from the metric and performing a separate variation with respect to these two fields \cite{Ferraris}.
The independent connection is assumed to describe the parallel transport and it is used to define the covariant derivative, while the metric tensor defines
distances and lengths. In the case
of General Relativity, the field equation obtained varying with respect to the {\em a priori} independent connection reduces to an algebraic equation
expressing the fact that the affine connection is rather the Levi-Civita connection of the metric itself. Consequently, the complete equivalence with 
the standard treatment of General Relativity through the metric formalism is restored.
However, such equivalence relies on 
the crucial assumption that the matter action is independent from the affine connection, that is the covariant derivatives contained in the Lagrangian 
density of the matter fields are the ones built from the metric connection. 

The simplest generalization that one can perform is to assume that the genuine covariant derivative is the one defined, in the most natural way, 
from the independent connection; then, the matter action turns out to be directly dependent from the affine connection. The resulting theory, 
that is the prototypical version of a metric-affine theory of gravity, is formally known as the Einstein-Cartan-Sciama-Kibble (ECSK) 
theory \cite{S, K}. 
In ECSK theory, it is still possible to show
that the independent (and not necessarily symmetric) connection
can be algebraically eliminated in favour of the metric and its derivatives, plus matter fields, losing any dynamical feature \cite{Hehl:1976kj}. This 
means that the 
connection is not propagating in the spacetime, but it is indissolubly confined to make couplings inside matter configurations. The eventual signature of the 
theory can be recovered only at level of the modified matter content obtained in the right hand side of the metric field equation (once the connection 
is completely eliminated therein), and in a new spin-spin contact interaction term modifying the dynamics of the matter fields.

Interestingly, nothing prevents us from adding further scalar invariants with the same dimension of the Ricci scalar in the Einstein-Cartan action.  
Einstein-Hilbert action is the only diffeo-invariant action leading to field equations of the second order and such that the connection is the metric one.
However, the
freedom obtained by introducing the independent connection as a new variable motivates a further generalization of ECSK theory.  
The action describing a metric-affine theory is assumed to be a suitable limit, at a certain order, of a some fundamental theory; one can then follow 
the standard 
approach of effective field theory, and consider all the operators having the dimension of the specific order of the approximation. In the present case 
of ECSK, these operators are all the possible second order (namely, all the operators of the same dimension of the scalar curvature) invariants that can
be formulated starting from the structures (torsion and nonmetricity) induced by the non-Riemannian nature of the actual spacetime.

Generalized metric-affine theories of gravity have been recently studied from different perspectives. Apart from the huge amount of work done in 
Einstein-Cartan
gravity (see for example \cite{Hehl:2007bn, Trautman:2006fp} for a bird's-eye view on recent aspects), progress was also made for
what concerns the higher order versions of these theories \cite{hehl1995}, also attracted by the tantalizing possibility of relaxing the theoretical puzzles associated 
with the recent discovery of accelerated expansion.
It is the case, for example, of Poincar\'{e} gauge theories of gravity \cite{Baekler:2010fr}, namely models whose action accommodates terms quadratic in 
torsion and curvature. The issue of the metric-affine counterpart for $f(\mathcal{R})$ modified theories of gravity was addressed in 
\cite{Sotiriou:2006qn} and, with a specific focus to the cosmological consequences, in \cite{Capozziello:2007tj}.

Different mechanisms were also proposed to alleviate the dark matter problem. It was suggested that a matter-antimatter asymmetry generated by the 
Hehl-Datta cubic term \cite{Hehl:1971qi} in the Dirac equation modified by the presence of torsion, could be at the origin of the presence of dark 
matter in the universe \cite{Poplawski:2011xf}. The decomposition of a metric-affine theory in scalars, vectors and tensors contribution, 
in a way to match the results of TeVeS phenomenology, was analyzed in \cite{Karahan:2011zq}. The authors of \cite{Tilquin:2011bu}, instead, 
suggested to fit Supernovae data replacing the dark matter content expected as from the $\Lambda$CDM model with a mean spin density of baryonic matter. 

The dynamical content of a metric-affine theory of gravity  was already analyzed in \cite{Vitagliano:2010sr} including the set of second order 
corrections built with torsion tensor. In this paper we are going to extend the previous result to the most general second order metric-affine theory of 
gravity, including quadratic 
operators in nonmetricity and torsion \cite{percacci}. Note that, as it will be showed on the basis of a dimensional analysis, this will not include 
the quadratic Ricci terms, whose presence would introduce \textit{brevi manu} new dynamical degrees of freedom. 
After a brief introduction of our notation, we study the conditions under which the independent connection
is reduced to the role of an auxiliary field in this metric-affine setup. 
% We trace back the consequences of the lack of dynamics of the connection, and
% show how this modifies the metric field equations, leading to an effective stress-energy tensor that carries the information of the extra matter 
% interactions. 
We then address the problem of the minimal requirements to make the connection 
dynamical. Higher order theories are taken into account, bringing particular attention to the very special case of $f(\mathcal{R})$ metric-affine 
theories. A final section is summarizing our results.

% 
% \textit{ Some specific peculiar cases are presented: first, we consider the case of fermionic matter, than we take into account a toy model
% of an intrinsically spinning fluid (also known as Weissenhoff fluid).} 

\section{Notation and Conventions}

Due to some ambiguities that can be found in the literature, and due to
the delicate role played by the symmetries of the tensorial objects hereafter considered, let us recall the basic definitions and 
conventions 
that we will use throughout the paper\footnote{It is worth stressing that some of the conventions used here are different from the ones adopted in 
\cite{Vitagliano:2010sr}.}. The covariant derivative with respect to the independent connection $\Gamma^\gamma{}_{\alpha\beta}$ of a generic $(1,1)$-tensor 
is defined as the following
\be
\s{\Gamma}{\nabla}_\alpha A^\beta{}_\gamma= \partial_\alpha A^{\beta}{}_{\gamma}{}+ \Gamma^{\beta}{}_{\alpha\delta} A^{\delta}{}_{\gamma}
-\Gamma^{\delta}{}_{\alpha\gamma} A^{\beta}{}_{\delta}\,. 
\ee 
Note that the connection $\Gamma^\gamma{}_{\alpha\beta}$ is not supposed to carry any symmetry. In particular, it is not supposed to covariantly 
conserve the metric. Using a combination of covariant derivatives of the metric, it is quite easy to show that the connection can be decomposed as
\be\label{Gamma}
\Gamma^{\gamma}{}_{\alpha\beta}=\big\{^{\,\gamma}_{\alpha\beta}\big\} +\frac{1}{2} \l(-Q_{\alpha}{}^{\gamma}{}_{\beta} +
Q^{\gamma}{}_{\beta\alpha} - Q_{\beta\alpha}{}^{\gamma}\r) + S_{\alpha\beta}{}^{\gamma}  - S_\beta{}^{\gamma}{}_{\alpha}+ S^{\gamma}{}_{\alpha\beta}\,,
\ee
where $\big\{^{\,\gamma}_{\alpha\beta}\big\}$ is the Levi-Civita connection of the metric (aka Christoffel symbols of the metric) and we have defined 
the nonmetricity tensor $Q_{\alpha\beta\gamma}\equiv\s{\Gamma}{\nabla}_\alpha g_{\beta\gamma}$ and the antisymmetric part of the 
connection, otherwise known as the Cartan torsion tensor, $S_{\alpha\beta}{}^{\gamma}\equiv \Gamma^{\gamma}{}_{[\alpha\beta]}$.
The curvature tensor associated with the connection $\Gamma^\gamma{}_{\alpha\beta}$ is defined by
\be
\R_{\mu\nu\rho}{}^\sigma=\partial_\nu\Gamma^{\sigma}{}_{\mu\rho}-\partial_\mu\Gamma^{\sigma}{}_{\nu\rho}
+\Gamma^{\alpha}{}_{\mu\rho}\Gamma^{\sigma}{}_{\nu\alpha}-\Gamma^{\alpha}{}_{\nu\rho} \Gamma^{\sigma}{}_{\mu\alpha}\,.
\ee
Here, the asymmetry of the connection deprives the curvature tensor of the usual properties of symmetry encountered in the (metric) Riemann tensor, 
that is 
the curvature tensor built from the Christoffels symbols of the metric. The only symmetry kept in the previous definition of the curvature tensor 
is the antisymmetry
with respect to the first two indices. This circumstance raises an ambiguity in the definition of the Ricci tensor of $\R_{\mu\nu\rho}{}^\sigma$: the 
usual prescription of the contraction of the second and the fourth indices provides the expression  
\be
\R_{\alpha\gamma\beta}{}^\gamma\equiv\R_{\alpha\beta}=\partial_\gamma \Gamma ^{\gamma}{}_{\alpha\beta}-\partial_\alpha\Gamma^{\gamma}{}_{\gamma\beta}
 +\Gamma ^{\gamma}{}_{\gamma\delta}\Gamma ^{\delta}{}_{\alpha\beta}-\Gamma^{\gamma}{}_{\alpha\delta} \Gamma ^{\delta}{}_{\gamma\beta}\,,
\ee
whose related Ricci scalar is $\R=g^{\alpha\beta}\R_{\alpha\beta}$. On the other hand, two other possible contractions can be performed; contracting the 
third and fourth indices we obtain $\R_{\alpha\beta\gamma}{}^\gamma\equiv\widetilde{\R}_{\alpha\beta}$, a tensor also known as the homothetic curvature,
antisymmetric with respect to its two indices, that reduces to $\R_{[\alpha\beta]}$ in the case of a symmetric (but still metric-incompatible) 
connection. The last possibility involves
the use of the metric tensor, $g^{\epsilon\gamma}g_{\delta\beta}\R_{\alpha\epsilon\gamma}{}^\delta\equiv{\hat{\R}}_{\alpha\beta}$. However, a further 
contraction with the metric of the two alternative definitions of the Ricci tensor gives $g^{\alpha\beta}\widetilde{\R}_{\alpha\beta}=0$ (due to the 
antisymmetry of the homothetic curvature) and $g^{\alpha\beta}\hat{\R}_{\alpha\beta}=-\R$, leaving the Ricci scalar uniquely determined.
In our effective field theory approach, we will mainly consider corrections to Einstein-Cartan Lagrangian of the same dimension of the Ricci scalar 
$\R$, so that the issue of the ambiguity will not be relevant for our purposes.

\section{The general action}

With what has been said till now, we are ready to provide a Lagrangian formulation of a metric-affine theory of gravity. Let 
us underline one more time the peculiar characteristic of metric-affine theories, namely, the possibility of a direct coupling of matter with the 
connection $\Gamma^\gamma{}_{\alpha\beta}$. The general action will be of the form
\bea
\mathcal{S}=\int d^4x \det [\mathcal{L}_G(g_{\mu\nu}, \Gamma^\gamma{}_{\alpha\beta})+\mathcal{L}_M(g_{\mu\nu}, \Gamma^\gamma{}_{\alpha\beta}, \psi)]\,,
\eea
where $g$ is the determinant of the metric, $\mathcal{L}_G(g_{\mu\nu}, \Gamma^\gamma{}_{\alpha\beta})$ and 
$\mathcal{L}_M(g_{\mu\nu}, \Gamma^\gamma{}_{\alpha\beta}, \psi)$ are respectively the gravitational and the matter Lagrangian density (where we have
made explicit the dependence of the matter from the connection), and $\psi$ is a convenient way to refer collectively to the matter fields included 
in the theory under scrutiny. 

We want to study the dynamics of the most general lowest order theory associated with the gravitational Lagrangian 
$\mathcal{L}_G(g_{\mu\nu}, \Gamma^\gamma{}_{\alpha\beta})$. Our prescription for constructing such general action is based on power counting of 
the dimensions of the gravitational terms. 
In natural units, where $c=\hbar=\kappa_B=1$, lengths and times have the same dimension; given the adimensionality of the
components of the metric tensor, the connection will have the dimension of the inverse of a length and consequently the Ricci tensor will have dimension 
$[\textrm{length}]^{-2}$. Since in natural units the total action must be dimensionless, the Lagrangian density must have dimension 
$[\textrm{length}]^{-4}$.
%that means that any term added to the action should appear with an overall constant with the suitable dimension
We can think the Lagrangian density as the product of a geometrical scalar invariant times an opportune overall 
constant, in the form of a power of a length $L_P$, to adjust the total dimension. For the Einstein-Cartan action, for example, the correct 
Lagrangian density is $\mathcal{L}_G^{EC}=\mathcal{R}/(16\pi L_P^2)$.

Discarding a cosmological constant term (that in principle can be included, but, since it is not playing any role in what follows, we are omitting here
for the sake of simplicity), it is not possible to build a Lagrangian density whose geometrical factor has dimension 
$[\textrm{length}]^{-1}$. This is trivially seen for the Ricci tensor case that is already of dimension $[\textrm{length}]^{-2}$. 
Nonmetricity and torsion tensor are also of dimension $[\textrm{length}]^{-1}$, but it is not possible to form a scalar
invariant from just one rank-three tensor saturated with the metric (rank two); for such reason they will appear only as quadratic terms, 
hence only as terms of order $[\textrm{length}]^{-2}$, that is at the same order of the Ricci scalar $\R$.

The most general gravitational action with a Lagrangian density of dimension $[\textrm{length}]^{-2}$ has the form  \cite{percacci}
\bea\label{most}
\hspace{-1cm}\mathcal{S}_G=\frac{1}{16\pi L_P^2}\int d^4x \det\l(\mathcal{R}+\sum_i a_i Q_{(i)}^2+\sum_i b_i Q_{(i)}* S_{(i)}+\sum_i c_i S_{(i)}^2\r)\,,
\eea
where the last three terms are a symbolic representation of all the possible independent contractions that can be obtained from nonmetricity $Q$ 
and torsion tensor $S$ (the symbol ``*'' in (\ref{most}) indicates the tensorial product between torsion and nonmetricity).
Using the symmetries of these two ($Q$ is symmetric with respect to its last two indices, while $S$ is antisymmetric with respect to its first two
indices), we find the following resulting combinations
\begin{itemize}
\item Pure nonmetricity terms
 
The way we can provide invariant quantities from the product of two nonmetricity tensors is through the saturation with three metric 
tensors\footnote{In what follows, we use the symbol ``$\circ$'' instead of the explicit indices for metric tensors and identity tensor in order 
to denote all the possible permutations of indices needed to saturate into a scalar the product of two tensors.} 
\be
Q_{\lambda\mu\nu}*Q_{\gamma\alpha\beta}*g^{\circ\circ}*g^{\circ\circ}*g^{\circ\circ}\,,
\ee
where the possible different independent contractions (having in mind the symmetry with respect to $\{\mu\nu\}$ and $\{\alpha\beta\}$) are the following
\be\label{1}
Q_{\lambda\mu\nu}*Q_{\gamma\alpha\beta}*g^{\lambda\gamma}*g^{\alpha\beta}*g^{\mu\nu}=Q_{\lambda\alpha}{}^{\alpha}Q^{\lambda}{}_{\sigma}{}^{\sigma}\,,
\ee
\be\label{2}
Q_{\lambda\mu\nu}*Q_{\gamma\alpha\beta}*g^{\lambda\mu}*g^{\gamma\alpha}*g^{\nu\beta}=Q^{\mu}_{\;\;\mu\nu}Q^{\alpha\;\;\nu}_{\;\;\alpha}\,,
\ee
\be\label{3}
Q_{\lambda\mu\nu}*Q_{\gamma\alpha\beta}*g^{\lambda\alpha}*g^{\gamma\mu}*g^{\nu\beta}=Q^{\mu}_{\;\;\alpha\beta}Q^{\alpha\;\;\beta}_{\;\;\mu}\,,
\ee
\be\label{4}
Q_{\lambda\mu\nu}*Q_{\gamma\alpha\beta}*g^{\lambda\gamma}*g^{\mu\alpha}*g^{\nu\beta}=Q_{\lambda\mu\nu}Q^{\lambda\mu\nu}\,,
\ee
\be\label{5}
Q_{\lambda\mu\nu}*Q_{\gamma\alpha\beta}*g^{\lambda\beta}*g^{\gamma\alpha}*g^{\mu\nu}=Q_{\lambda\alpha}{}^{\alpha}Q^{\sigma\;\;\lambda}_{\;\;\sigma}\,.
\ee

\item {Mixed terms}

The generic nonmetricity-torsion interaction term is written as
\be\label{generic}
Q_{\lambda\mu\nu}*S_{\alpha\beta}^{\;\;\;\;\gamma}*\delta^{\circ}{}_{\circ}*g^{\circ\circ}*g^{\circ\circ}\,.
\ee 
The identity tensor $\delta^{\circ}{}_{\circ}$ can be used either for contracting indices in the torsion tensor or to raise one of the indices of the 
nonmetricity.
In the former case, exploiting the antisymmetry of torsion in $\{\alpha\beta\}$, we can only obtain the object 
$S_{\alpha}\equiv\delta_{\gamma}^{\;\;\beta}S_{\alpha\beta}^{\;\;\;\;\gamma}$ 
(or an equivalent result contracting $\alpha$ and $\gamma$), from which we can construct the following combinations
\be\label{first}
Q_{\lambda\mu\nu}*S_{\alpha}*g^{\lambda\mu}*g^{\nu\alpha}\,,
\ee
\be\label{fasullo}
Q_{\lambda\mu\nu}*S_{\alpha}*g^{\lambda\nu}*g^{\mu\alpha}\,,
\ee
\be\label{second}
Q_{\lambda\rho}{}^\rho*S_{\alpha}*g^{\lambda\alpha}\,,
\ee
where (\ref{first}) and (\ref{fasullo}) are two proportional terms, due to the symmetry of $Q$ in $\{\mu\nu\}$.
If $\delta^{\circ}{}_{\circ}$ in (\ref{generic}) acts on the nonmetricity, we have a double choice, 
either $\delta^{\lambda}_{\;\;\gamma}Q_{\lambda\mu\nu}$ or $\delta^{\mu}_{\;\;\gamma}Q_{\lambda\mu\nu}$. In the first case we have
\bea
\hspace{-2.5cm}Q_{\gamma\mu\nu}S_{\alpha\beta}^{\;\;\;\;\gamma}g^{\mu\alpha}g^{\nu\beta}=
-Q_{\gamma\mu\nu}S_{\beta\alpha}^{\;\;\;\;\gamma}g^{\mu\alpha}g^{\nu\beta}=
-Q_{\gamma\mu\nu}S_{\alpha\beta}^{\;\;\;\;\gamma}g^{\mu\beta}g^{\nu\alpha}=
-Q_{\gamma\nu\mu}S_{\alpha\beta}^{\;\;\;\;\gamma}g^{\mu\beta}g^{\nu\alpha}=\nn\\
=-Q_{\gamma\mu\nu}S_{\alpha\beta}^{\;\;\;\;\gamma}g^{\nu\beta}g^{\mu\alpha}=0\,,
\eea
in the second situation we have
\be\label{third}
Q_{\lambda\gamma\nu}*S_{\alpha\beta}^{\;\;\;\;\gamma}*g^{\alpha\lambda}*g^{\beta\nu}\,.
\ee
Summarizing, the three independent terms one can construct from mixed terms are (\ref{first}), (\ref{second}) and (\ref{third})

\item {Pure torsion terms}

In this last case, the scalar quantity is built saturating the product of two torsion tensors with two $\delta$'s and one metric tensor 
in a structure like the following
\be
S_{\mu\nu}^{\;\;\;\;\rho}*S_{\alpha\beta}^{\;\;\;\;\gamma}*\delta^{\circ}{}_{\circ}*\delta^{\circ}{}_{\circ}*g^{\circ\circ}\,,
\ee 
whose possible combinations assume the form
\bea
g^{\mu\nu}S_{\mu}S_{\nu}\,,\\
g^{\mu\nu}S_{\mu\lambda}^{\;\;\;\;\sigma}S_{\nu\sigma}^{\;\;\;\;\lambda}\,,\\
g^{\mu\alpha}g^{\nu\beta}g_{\lambda\gamma}S_{\mu\nu}^{\;\;\;\;\lambda}S_{\alpha\beta}^{\;\;\;\;\gamma}\,.
\eea
\end{itemize}

Taking into account the independent terms identified above, we can rewrite the three last sums in (\ref{most}) 
\bea\label{most+}
\hspace{-2.1cm}\sum_i a_i Q_{(i)}^2= a_1 Q_{\lambda\mu}{}^{\mu} Q^{\lambda}{}_{\nu}{}^{\nu}+ a_2 Q_{\mu}{}^{\mu\nu} Q_{\lambda}{}^{\lambda}{}_{\nu}
+a_3 Q_{\mu}{}^{\lambda\nu}Q_{\lambda}{}^{\mu}{}_{\nu}+a_4 Q_{\lambda\mu\nu}Q^{\lambda\mu\nu}+a_5 Q_{\nu}{}^{\nu\lambda} Q_{\lambda\mu}{}^{\mu},\nn\\
\hspace{-2.1cm}\sum_i b_i Q_{(i)}*S_{(i)}=b_1 Q_{\lambda}{}^{\lambda}{}_{\mu}S^{\mu}{}_{\nu}{}^{\nu}+b_2 Q_{\lambda\mu}{}^{\mu}S^{\lambda}{}_{\nu}{}^{\nu}
+ b_3 Q^{\lambda}{}_{\mu}{}^{\nu} S_{\lambda\nu}{}^{\mu}\,,\nn\\
\hspace{-2.1cm}\sum_i c_i S_{(i)}^2= c_1 S_{\mu\nu}{}^{\nu} S^{\mu}{}_{\lambda}{}^{\lambda} + c_2 S_{\mu\lambda}{}^{\nu} S^{\mu}{}_{\nu}{}^{\lambda}
+ c_3 S_{\mu\nu}{}^{\lambda} S^{\mu\nu}{}_{\lambda}\,.
\eea

It is interesting to note that the full Lagrangian density is free of terms obtained by the covariant derivative
of nonmetricity and torsion. The reason is easily understood. The independent connection can be decomposed, as already shown in (\ref{Gamma}), in the 
Levi-Civita connection of the metric plus a combination $K$ of terms in $Q$ and $S$, called distortion tensor
\be\label{distortion}
K_{\alpha\beta}{}^{\gamma}=\frac{1}{2} \l(-Q_{\alpha}{}^{\gamma}{}_{\beta} +
Q^{\gamma}{}_{\beta\alpha} - Q_{\beta\alpha}{}^{\gamma}\r) + S_{\alpha\beta}{}^{\gamma}  - S_\beta{}^{\gamma}{}_{\alpha}+ S^{\gamma}{}_{\alpha\beta}\,.
\ee
Using this last condition, one can think to write the covariant derivative with respect to the independent $\Gamma$ as a Riemannian part 
(namely a covariant derivative with respect to the metric) plus another one encoding the non-Riemannian structures of the spacetime 
\bea
\s{\Gamma}{\nabla}_\mu Q_{\alpha\beta\gamma}=\s{\{\}}{\nabla}_\mu Q_{\alpha\beta\gamma}+\textrm{contractions $K*Q$}\,,\nn\\
\s{\Gamma}{\nabla}_\mu S_{\alpha\beta}^{\4\gamma}=\s{\{\}}{\nabla}_\mu S_{\alpha\beta}^{\4\gamma}+\textrm{contractions $K*S$}\,;
\eea
the first term on the right hand side of both previous equations is a total divergence resulting in a surface term, the second terms are instead 
contractions of the distortion tensor with $Q$ and $S$, i.e. (quadratic) combinations of torsion and nonmetricity, already included in the 
classification considered above.

It should be noted that one could still in principle enrich the gravitational Lagrangian with yet another 
term, proportional to the Ricci scalar built from the metric tensor. This is the only second order term that can be built from the metric. 
This Ricci scalar can be expressed in terms of the Ricci scalar of the connection plus a combination of
nonmetricity and torsion tensors \cite{Faria:2013moa}
\be\label{referee}
\hspace{-1.5cm}\frac{1}{16\pi L_P^2}\int d^4x \det R=\frac{1}{16\pi L_P^2}\int d^4x \det[\mathcal{R}-(K^\alpha{}_{\rho\alpha} K^\rho{}_{\mu}{}^{\mu}
+K^\alpha{}_{\rho\mu} K^{\rho\mu}{}_{\alpha})]\,,
\ee
(we have discarded total derivatives leading to surface terms) and can be easily reabsorbed in the terms already present in the Lagrangian density. 
It is worth noticing here that such a procedure is valid 
only in the case of general second order Lagrangian. For higher order modified gravity models, it is not true anymore, see for example the 
$R+f(\mathcal{R})$ theory \cite{Harko:2011nh} that leads to the introduction of an effective further degree of freedom. 

In order to conclude this section, a comment is still due: the eleven quadratic terms in torsion and nonmetricity in (\ref{most+}) should 
be in principle complemented with the parity-violating terms obtained from the contraction of curvature tensor, torsion and
nonmetricity (as well as for eventual combinations of them) with the totally
antisymmetric Levi-Civita tensor $\epsilon_{\alpha\beta\gamma\delta}$. Anyway, in this paper we are working with the simplifying
(realistic) hypothesis of a parity preserving Lagrangian (as already assumed by \cite{macias}); such a requirement has also another root: in the same
spirit of \cite{Sotiriou:2006qn, Vitagliano:2010sr, Hehl:1981ed} the Lagrangian can be further required to fulfill a minimal coupling inspired
construction, i.e., the general action should include only invariants that can be
built using the metric tensor to contract indexes. This procedure is automatically 
selecting the parity even terms in the action, as taken into account in (\ref{most+}).

\section{The dynamical content of the independent connection}

We are basically interested in the field equation for the connection, since it is the one that should give us the possibility to re-express it
as a function of matter fields plus Christoffel symbols of the metric.

Let us derive the variation of nonmetricity and torsion. One has
\bea
\delta Q_{\alpha\beta\gamma}=\delta\s{\Gamma}{\nabla}_\alpha g_{\beta\gamma}=\partial_\alpha \delta g_{\beta\gamma}
+2g_{\lambda\mu}\Gamma^\lambda_{\alpha(\beta}g_{\gamma)\nu}\delta g^{\mu\nu}
-2g_{\lambda(\gamma}\delta^\nu_{\beta)}\delta^\mu_\alpha\,\delta\Gamma_{\mu\nu}^{\lambda}\,,\\
\delta S_{\alpha\beta}^{\4\lambda}=\delta_\alpha^{[\mu}\delta_\beta^{\nu]}\delta\Gamma^\lambda_{\mu\nu}\,.
\eea
% I'm interested just in the variation wrt the connection, so the first term just enter in the 1st field equation. The variation wrt connection gives
% \be
% -g_{\sigma\gamma}\delta\Gamma_{\beta\alpha}^{\sigma} - g_{\beta\sigma}\delta \Gamma_{\gamma\alpha}^{\sigma}
% \ee
% this quantity must be multiplied for an opportune combination of metric tensors. Please note that, when doing the variation, we are varying wrt to the
% connection $\Gamma_{\mu\nu}^{\lambda}$, so the indices must be fixed properly.
% \be
% \delta Q_{\delta\epsilon\rho}\rightarrow
% \delta \Gamma_{\mu\nu}^{\lambda}(-g_{\lambda\rho}\delta^{\mu}_{\epsilon}\delta^{\nu}_{\delta}-g_{\epsilon\lambda}\delta^{\mu}_{\rho}\delta^{\nu}_{\delta})
% \ee
% \be
% \delta Q_{\gamma\alpha\beta}\rightarrow
% \delta \Gamma_{\mu\nu}^{\lambda}(-g_{\lambda\beta}\delta^{\mu}_{\alpha}\delta^{\nu}_{\gamma}-g_{\alpha\lambda}\delta^{\mu}_{\beta}\delta^{\nu}_{\gamma})
% \ee
% The terms from (\ref{1}) to (\ref{5}) have the same structure, that we can write in the following form
% \bea
% \delta(Q_1*Q_2*f(g's))=f(g's)(Q_1\delta Q_2+Q_2\delta Q_1)
% \eea

Without loss of generality, the global structure of the Palatini field equation will be of the form
\be\label{general}
\hspace{-2cm}\frac{1}{\det}[\s{\Gamma}{\nabla}_\lambda (\det g^{\mu\nu})-\s{\Gamma}{\nabla}_\sigma(\det g^{\sigma\mu})\delta^\nu_\lambda]+\sum_j\alpha_j\,Q^{(j)}
+\sum_j \beta_j\,S^{(j)}
=(8\pi L_P^2)\Delta^{\mu\nu}_{\;\;\;\;\lambda}\,,
\ee
where 
\bea
\sum_j\alpha_j\,Q^{(j)}&=&\a1 \delta^{\mu}{}_{\lambda} Q^{\nu}{}_{\alpha}{}^{\alpha} +\a2 g^{\mu\nu} Q_{\alpha}{}^{\alpha}{}_{\lambda}
+ \a3 \delta^{\nu}{}_{\lambda} Q^{\alpha}{}_{\alpha}{}^{\mu} + \a4 
Q^{\mu\nu}{}_{\lambda}+\nn\\
&&+\a5 
Q_{\lambda}{}^{\nu\mu} +  \a6 
Q^{\nu\mu}{}_{\lambda} 
+ \a7 \delta^{\mu}{}_{\lambda} Q^{\alpha}{}_{\alpha}{}^{\nu} + \a8 g^{\mu\nu} Q_{\lambda\alpha}{}^{\alpha}+\nn\\
&&+ \a9 \delta^{\nu}{}_{\lambda} Q^{\mu}{}_{\alpha}{}^{\alpha}\,,
\eea
\bea
\sum_j \beta_j\,S^{(j)}&=&\b1 g^{\mu\nu} S_{\lambda\alpha}{}^{\alpha} + \b2 \delta^{\nu}{}_{\lambda} 
S^{\mu}{}_{\alpha}{}^{\alpha} + \b3 \delta^{\mu}{}_{\lambda} 
S^{\nu}{}_{\alpha}{}^{\alpha} + \b4 S^{\nu}{}_{\lambda}{}^{\mu} +\nn\\
&&+\b5 S^{\nu\mu}{}_{\lambda} + \b6 S^{\mu}{}_{\lambda}{}^{\nu}\,,
\eea
and $\alpha_j$ and $\beta_j$ are some linear combinations of the primitive coefficients $a_i$, $b_i$ and $c_i$ used in equation (\ref{most}), 
\bea
\hspace{-2.5cm}\{\alpha_j\}\equiv&&\{\frac{2a_5+b_2-1}{2};\;2a_2;\frac{4a_5-b_1}{2};\;\frac{8a_4-b_3}{2};\;2a_3-1;\frac{4a_3+b_3}{2};\;
\frac{4a_2+b_1+2}{2};\nn\\
\hspace{-2.5cm}&&\frac{2a_5+1}{2};\;\frac{8a_1-b_2}{2}\}\,,\nn\\
\hspace{-2.5cm}\{\beta_j\}\equiv&&\{b_1+2;\;2b_2-c_1;\;b_1+c_1-2;\;c_2+2;\;2c_3-b_3;\;b_3-c_2\}\,.
\eea
The right hand side of (\ref{general}) is given by the so-called hypermomentum tensor 
$\Delta^{\mu\nu}{}_\lambda\equiv-\frac{2}{\det}\frac{\delta S_M (g,\Gamma,\psi)}{\delta\Gamma^\lambda{}_{\mu\nu}}$, that is, the tensor describing the 
intrinsic properties of matter as spin angular momentum, shear and dilation current \cite{Hehl:1976kv}. Also note that
\bea
\hspace{-2.5cm}\frac{1}{\det}[\s{\Gamma}{\nabla}_\lambda (\det g^{\mu\nu})-\s{\Gamma}{\nabla}_\sigma(\det g^{\sigma\mu})\delta^\nu_\lambda]=
\delta^{\nu}{}_{\lambda} Q_{\alpha}{}^{\alpha\mu} + \frac{1}{2} g^{\mu\nu} Q_{\lambda\alpha}{}^{\alpha} - Q_{\lambda}{}^{\mu\nu} 
-  \frac{1}{2} \delta^{\nu}{}_{\lambda} Q^{\mu}{}_{\beta}{}^{\beta}\,,\nn\\
\eea
that is, we can reformulate the Palatini equation just in terms of an expression linear in the torsion tensor and in the nonmetricity tensor, 
that we can symbolically rewrite as 
\be\label{Pal}
\sum_j\ta{j}\,Q^{(j)}+\sum_j \beta_j\,S^{(j)}=(8\pi L_P^2)\Delta^{\mu\nu}_{\;\;\;\;\lambda}\,,
\ee
with $\ta3=\a3+1$, $\ta5=\a5-1$, $\ta8=\a8+\frac{1}{2}$, $\ta9=\a9-\frac{1}{2}$ and $\ta{j\neq3,5,8,9}=\a{j}$. 
We can contract this equation in three different independent ways: with the metric $g_{\mu\nu}$, with $\delta_\mu^\lambda$ and with $\delta_\nu^\lambda$.
We obtain respectively
\bea
(4\ta2 +\ta3 +\ta4 +\ta6 +\ta7) Q^{\alpha}{}_{\lambda\alpha} + (\ta1 +\ta5 + 4\ta8 +\ta9) 
Q_{\lambda}{}^{\alpha}{}_{\alpha}+\nn\\
\hspace{2.5cm}+ (4\b1 +\b2 +\b3 - \b4 - \b6) S_{\lambda}{}^{\alpha}{}_{\alpha}=(8\pi L_P^2)\Delta_{\alpha}{}^{\alpha}{}_{\lambda}\,,
\eea
\bea
(\ta2 +\ta3 +\ta4 +\ta5 + 4\ta7) Q^{\alpha\nu}{}_{\alpha} + (4\ta1 +\ta6 +\ta8 +\ta9) 
Q^{\nu\alpha}{}_{\alpha}+\nn\\
\hspace{2.5cm} + (\b1 +\b2 + 4\b3 +\b4 +\b5) S^{\nu\alpha}{}_{\alpha}=(8\pi L_P^2)\Delta^{\alpha\nu}{}_{\alpha}\,,
\eea
\bea
(\ta2 + 4\ta3 +\ta5 +\ta6 +\ta7) Q^{\alpha\mu}{}_{\alpha} + (\ta1 +\ta4 +\ta8 + 4\ta9) 
Q^{\mu\alpha}{}_{\alpha}+\nn\\
\hspace{2.5cm} + (\b1 + 4\b2 +\b3 - \b5 +\b6) S^{\mu\alpha}{}_{\alpha}=(8\pi L_P^2)\Delta^{\mu\alpha}{}_{\alpha}\,.
\eea
This is a simple linear system whose solution can be written as follows
\bea
Q^{\alpha\rho}{}_{\alpha}=(8\pi L_P^2)(A_1 \Delta_{\alpha}{}^{\alpha\rho}+B_1 \Delta^{\alpha\rho}{}_{\alpha}+C_1\Delta^{\rho\alpha}{}_{\alpha})\,,\nn\\
Q^{\rho\alpha}{}_{\alpha}=(8\pi L_P^2)(A_2 \Delta_{\alpha}{}^{\alpha\rho}+B_2 \Delta^{\alpha\rho}{}_{\alpha}+C_2\Delta^{\rho\alpha}{}_{\alpha})\,,\nn\\
S^{\rho\alpha}{}_{\alpha}=(8\pi L_P^2)(A_3 \Delta_{\alpha}{}^{\alpha\rho}+B_3 \Delta^{\alpha\rho}{}_{\alpha}+C_3\Delta^{\rho\alpha}{}_{\alpha})\,.
\eea
where $A_i$, $B_i$ and $C_i$ are some elementary, but rather lenghty, expressions of the coefficients $\ta{i}$ and $\b{i}$.
We can now use these three equations to substitute the corresponding terms in (\ref{Pal}); their contribution is fully determined by the matter content 
of the theory, so we can move them on the right hand side, where they are collectively denoted as ``$[f(\textrm{traces of }\Delta)]^{\mu\nu}{}_\lambda$''. 
What remains is the equation
\bea\label{master}
\hspace{-1cm}\ta4 Q^{\mu\nu}{}_{\lambda}+\ta5 Q_{\lambda}{}^{\nu\mu} +\ta6 Q^{\nu\mu}{}_{\lambda} + \b4 S^{\nu}{}_{\lambda}{}^{\mu} 
+\b5 S^{\nu\mu}{}_{\lambda} + \b6 S^{\mu}{}_{\lambda}{}^{\nu}=\nn\\
\hspace{4cm}=(8\pi L_P^2)\Delta^{\mu\nu}_{\;\;\;\;\lambda}
+[f(\textrm{traces of }\Delta)]^{\mu\nu}{}_\lambda.
\eea
The antisymmetric part with respect to $\{\mu\nu\}$ pairs of indices of the previous equation gives an equation to express the torsion tensor in terms of
the antisymmetric part of nonmetricity tensor plus terms in hypermomenta
\be
\hspace{-2.4cm}\b5 S^{\nu\mu}{}_{\lambda} + (\b6-\b4) S^{[\mu}{}_{\lambda}{}^{\nu]}\!=\!
(\ta6-\ta4) Q^{[\mu\nu]}{}_{\lambda}+(8\pi L_P^2)\Delta^{[\mu\nu]}_{\;\;\;\;\lambda}+[f(\textrm{traces of }\Delta)]^{\mu\nu}{}_\lambda
\equiv\Theta^{\mu\nu}{}_\lambda\,,
\ee
that can be solved considering a suitable combination of the three different permutations of the indices $(\mu\nu\lambda)\rightarrow(\lambda\mu\nu)$ and 
$(\mu\nu\lambda)\rightarrow(\lambda\nu\mu)$.
At the end, we get:
\bea\label{torsione}
\hspace{-1.2cm}S_{\mu\nu\lambda}&=& \frac{2 \b5 \Theta_{\mu\nu\lambda}-\b4 (\Theta_{\mu\nu\lambda}-\Theta_{\lambda\mu\nu}+\Theta_{\lambda\nu\mu})
+\b6 (\Theta_{\mu\nu\lambda}-\Theta_{\lambda\mu\nu}+\Theta_{\lambda\nu\mu})}{(\b4+2 \b5-\b6) (\b4-\b5-\b6)}\nn\\
&&=[\widehat{f}(\Delta)]_{[\mu\nu]\lambda}-\frac{2(\ta6-\ta4)}{\b4+2\b5-\b6}Q_{[\mu\nu]\lambda}\,.
\eea
Since we are not interested in the exact form of the contribution of matter to the torsion tensor, we have 
here defined another tensor $[\widehat{f}(\Delta)]_{\mu\nu\lambda}$ that includes all the contributions coming from the hypermomenta in 
$\Theta_{\mu\nu\lambda}$.
Note that the tensor $\Theta_{\mu\nu\lambda}$, and hence the torsion tensor $S_{\mu\nu\lambda}$, is linear in nonmetricity $Q_{\mu\nu\lambda}$.
Using this expression in equation (\ref{master}) to eliminate the torsion, we can rewrite it in the form
\be\label{nonmet}
\xi_1 Q_{\mu\nu\lambda}+\xi_2Q_{\lambda\nu\mu}+\xi_3Q_{\nu\mu\lambda}=[\overline{f}(\Delta)_{\mu\nu\lambda}]\,,
\ee
where $\xi_i \equiv \xi_i(\tilde{\alpha}_j,\beta_k)$ are some coefficients determined by the equations (\ref{master}) and (\ref{torsione}) and
$[\overline{f}(\Delta)]_{\mu\nu\lambda}$ is defined, in analogy to $f$ and $\hat{f}$, as the collective contribution from matter to the right hand side 
of the expression; equation (\ref{nonmet}) can be now solved with respect to $Q_{\mu\nu\lambda}$ adding and subtracting the further two 
equations obtained permuting the indices $(\mu\nu\lambda)\rightarrow(\lambda\mu\nu)$ and 
$(\mu\nu\lambda)\rightarrow(\nu\mu\lambda)$. Having expressed nonmetricity in terms of just matter fields, we can reuse it in the equation for torsion 
to have another expression using just the matter fields. At the end, the total connection, that can be written in terms of Christoffel symbols of the 
metric 
plus distortion (where we recall that the distortion tensor (\ref{distortion}) is a combination of nonmetricity and torsion), is hence reduced to a 
(not trivial) expression of 
metric with its derivatives and of matter fields under the guise of the hypermomenta combination. 

An important point to stress is the following. Given a matter Lagrangian that does
contain at most linear terms in the covariant derivative, and hence linear terms in the connection, our demonstration shows the lack of
dynamics, and the consequent reduction to an auxiliary field, of $\Gamma^{\lambda}{}_{\mu\nu}$. On the other hand, some specific and exotic forms of 
matter can evade this fulfillment. Anyway, equations of motion of matter fields are required to be at most of second order, which forces the 
matter Lagrangian to contain only linear derivatives. Therefore, even in the most convoluted case, the hypermomentum tensor will be algebraic in the 
connection and the latter can be still eliminated at the component level.

What is the physical consequence of the lack of dynamics of the connection? Once it has been shown that the connection can be algebraically written 
in terms of derivatives of the metric and matter fields, it is clear that we can substitute all the terms explicitly dependent on connection (or torsion, 
or nonmetricity) in the field equation obtained varying with respect to the metric. Due to the extreme length of the equation, we will omit to write 
it here completely. It is anyway clear that, varying the general action (\ref{most}) with respect to the metric, we will basically obtain terms that are 
quadratic in the connection (torsion/nonmetricity), and hence that will carry extra contributions of the kind
``(hypermomentum)$^2$'' to the effective stress energy tensor. This is similar to what happens in Einstein-Cartan theory as shown in \cite{Hehl:1976kj},
with the main difference that now, because of the presence of nonmetricity terms, the field equations will contain new terms of different nature, 
coupling matter fields (in the form of hypermomenta) to Christoffel symbols.

\section{A new role for matter fields}

Which kind of matter is sensitive to a minimal coupling with the connection through the covariant derivative? Clearly, scalar fields, having no spin, 
result to be neither sources for torsion nor for non-metricity, as can be easily understood from the fact that the covariant derivative of a scalar 
field is equal to its partial derivative and henceforth does not depend on the Levi-Civita independent connection. 

Gauge fields (like the electromagnetic field) also do not couple to the connection, since they can be covariantly defined in terms of exterior 
derivatives. As a consequence, photons are not affected by the presence of torsion and non-metricity, and the causality of the theory will be
preserved and completely determined from the metric structure of the spacetime.
In principle one can think to break the gauge symmetry by adding a term that is manifestly gauge non-invariant, as for the mass term in the Einstein-
Proca-Maxwell action, whose consequence is the emergence of a coupling between the spin of the particles and the non-Riemannian structures.

From the physical point of view, it would be interesting to study some concrete
examples of matter fields coupled to connection, namely those fields for which the
hypermomentum does not vanish. A (phenomenological) candidate to source connection dynamics could be a semiclassical
spinning dust matter distribution, {\em alias} a generalization of the perfect fluid in the case
of non-vanishing spin, a fluid otherwise dubbed Weyssenhoff fluid \cite{weiss}; even though
such kind of matter is an interesting toy model, it has an unsatisfactory theoretical
formulation, since there is no unambiguous Lagrangian formulation able to describe it. Instead,
one has to postulate some convective forms for the energy-momentum and spin-
angular momentum tensors, plus some restrictions to the fluid spin in order to
ensure the integrability conditions of the equations of motion of the particles.

A better motivated example of matter field sourcing the equation for the connection is the Dirac field or any massive
vector field or tensor field  (not necessarly in the domain of the standard model)
, whose actions contain explicit dependences on the covariant
derivative and hence lead to field equations with non zero hypermomenta. 
In those cases, the fields are potentially able to induce non-metricity
and/or torsion\footnote{It is worth reminding that those fields that do not introduce
either torsion or nonmetricity because not coupled to the connection, also will not be affected by torsion and nonmetricity
even if other matter fields produce it.}.
As a pedagogical example let us consider the minimally coupled Dirac Lagrangian
in the simplified context of Einstein-Cartan theory. Standard model Dirac fermions
can be coupled only to the metric-compatible part of the linear connection, so
the matter Lagrangian reads
\bea
\mathcal{L}_{{Dirac}}(\Gamma)&=&(\s{\Gamma}{\nabla}_\mu \bar{\psi}) \gamma^\mu \psi-\bar{\psi} \gamma^\mu (\s{\Gamma}{\nabla}_\mu \psi)-m\bar{\psi}\psi=
\nn\\
&&=\mathcal{L}_{{Dirac}}(\{\})-\bar{\psi}\gamma^{[\mu}\gamma^\nu\gamma^{\rho]}\psi K_{\rho\nu\mu}\,,
\eea
where the last line expresses the splitting of the Dirac Lagrangian into a Riemannian piece plus a non-Riemannian correction due to the
spin-torsion coupling of the spinors\footnote{Here the distortion tensor should be read as a combination of the sole torsion tensor, as can be found 
from (\ref{Gamma}) once that nonmetricity is set to zero.}. In this case the hypermomentum tensor
$\Delta^{\mu\nu\rho}=\bar{\psi}\gamma^{[\mu}\gamma^\nu\gamma^{\rho]}\psi$ is the new source appearing in the field equation for the 
independent connection. If the equation for the connection is algebraic, as for the ECSK theory, the spin-connection coupling encoded in the 
hypermomentum tensor will result (once the connection is eliminated from the metric field equation) in an effective correction to the 
stress energy tensor in the form of a spin-spin term.

\section{Higher orders terms}

It is an easy task to argue that scalar Lagrangian corrections of the order $[\textrm{length}]^{-2n}$, with integer $n$, are the only ones that can be 
written starting from our elementary geometrical objects. This is essentially due to the fact that the only quantities carrying dimension 
$[\textrm{length}]^{-1}$ are odd-rank objects (torsion, nonmetricity and covariant derivative), and they cannot be trivially saturated with the 
(even-rank) metric. For such reason, the next-order invariants are all the possible terms with dimension $[\textrm{length}]^{-4}$. It goes beyond the 
scope of this paper to enumerate all the possible invariants of the fourth order (just for comparison, it should be taken into account that in a 
spacetime with torsion and vanishing nonmetricity there are 151 independent scalar invariants \cite{chris}).
It is anyway possible to show that these terms are 
inevitably introducing further degrees of freedom, even assuming the simplifying hypothesis of matter fields not coupled to the connection.
It is what for example occurs for the generalized Palatini theories considered in \cite{Vitagliano:2010pq}, whose specific choice of a Lagrangian 
density of the form 
$\mathcal{R}+\mathcal{R}_{\mu\nu}\mathcal{R}_{\kappa\lambda}(ag^{\mu\kappa}g^{\nu\lambda}+bg^{\mu\lambda}g^{\nu\kappa})$ has been shown to be
equivalent to Einstein gravity plus a (dynamical) Proca vector field.

A rather peculiar case is the metric-affine version of $f(\mathcal{R})$ theories of gravity \cite{Sotiriou:2006qn, Sotiriou:2008rp}. The Ricci scalar 
$\mathcal{R}$ is invariant under the projective transformation
\bea
\Gamma_{\;\;\mu\nu}^{\rho} \rightarrow\Gamma_{\;\;\mu\nu}^{\rho}+\delta_{\;\;\mu}^{\rho}\xi_{\nu}\,,
\eea
where $\xi_\nu$ is an arbitrary covariant vector field. Consequently also any function of $\mathcal{R}$ will respect the same symmetry. 
While this issue is not a problem when matter does not couple to the connection, for a metric-affine theory this feature can lead to inconsistent
field equations. 
In general the matter Lagrangian is not projective invariant, neither it is reasonable to restrict the matter content to those fields that are 
fulfilling
this property. To circumvent the problem, it is necessary to break the projective invariance of the gravitational sector, fixing the four degrees of 
freedom of the transformation (related to the four components of the vector field $\xi_\nu$) by a Lagrange multiplier. 

Note that the projective 
invariance is automatically broken if distortion-squared terms and, {\em a fortiori}, higher order curvature invariants are added to the action\footnote{
For the latter case see for example \cite{hehl1995, percacci}.}; 
such theories have been shown already to carry further dynamics. The simplest example is the theory considered in \cite{Harko:2011nh}, 
where the Lagrangian density is assumed to be $\mathcal{L}_G = R + f_1(\mathcal{R})$ (here $R$ is the Ricci scalar built from the (metric) Riemann
tensor and $f_1(\mathcal{R})$ is a function of the Ricci scalar of the curvature tensor).  We can now use (\ref{referee}) to write 
$R = \mathcal{R} +(\textrm{distortion})^2$, 
it is then clear that the total Lagrangian turns out to be $\mathcal{L}_G = f_2(\mathcal{R})+(\textrm{distortion})^2$, where 
$f_2(\mathcal{R})=\mathcal{R}+ f_1(\mathcal{R})$.
However, this theory has been shown to carry one more effective degree of freedom. Here, we want to remain in the 
realm of $f(\mathcal{R})$ theories of gravity, so we will skip further discussions about that.
Since the number of degrees of freedom to be fixed is four, and since the projective transformation suggests that the goal of breaking 
projective invariance should be achieved by constraining the connection, it is reasonable to propose an additional term in the gravitational Lagrangian
involving a contraction of either nonmetricity tensor or torsion tensor.

As already shown in \cite{Sotiriou:2007yd}, the term of the form $A^\mu Q_\mu\equiv A^\mu g^{\alpha\beta}Q_{\mu\alpha\beta}$, previously proposed in 
\cite{Hehl:1981ed} is not suitable for a generic $f(\mathcal{R})$ 
metric-affine theories, since it requires the function of the Ricci scalar to reduce to the Einstein-Cartan term and the theory results to have no 
solutions of the field equations whenever the $f(\mathcal{R})$ is non-linear. Interestingly, the result is still valid even if we try to fix the
four degrees of freedom through the other independent contraction of the nonmetricity tensor, namely adding to the Lagrangian density the Lagrange 
multiplier the $B^\mu \widetilde{Q}_\mu\equiv B^\mu g^{\alpha\beta}Q_{\alpha\beta\mu}$, as it can be easily proved: in a torsion-less theory 
without matter fields, the independent Levi-Civita connection can be written 
from (\ref{Gamma}) as
\be\label{dim1}
\Gamma^{\gamma}{}_{\alpha\beta}=\big\{^{\,\gamma}_{\alpha\beta}\big\} +\frac{1}{2} \l(-Q_{\alpha}{}^{\gamma}{}_{\beta} +
Q^{\gamma}{}_{\beta\alpha} - Q_{\beta\alpha}{}^{\gamma}\r)\,;
\ee
on the other hand, the field equation of the connection reduces to the usual $f(\mathcal{R})$-Palatini equation\footnote{The condition about the absence
of matter fields implies also the vanishing of the Lagrange multipliers $A_\mu$ and
$B_\mu$ that can be proven to be proportional to the trace of the hypermomentum tensor.}, that can be solved with respect to the 
connection to give (see for example \cite{Sotiriou:2009xt})
\be\label{dim2}
\Gamma^{\gamma}{}_{\alpha\beta}=\big\{^{\,\gamma}_{\alpha\beta}\big\} +\frac{1}{2f'(\mathcal{R})}(2\partial_{(\alpha} f'(\mathcal{R}) 
\delta^\gamma_{\beta)}-g^{\gamma\sigma}g_{\alpha\beta}\partial_\sigma f'(\mathcal{R}))\,.
\ee
Equating (\ref{dim1}) and (\ref{dim2}) gives a condition expressing the contribution to nonmetricity coming from the gravitational sector of the theory
\be\label{dim3}
-Q_{\alpha}{}^{\gamma}{}_{\beta} +Q^{\gamma}{}_{\beta\alpha} - Q_{\beta\alpha}{}^{\gamma}=
\frac{1}{f'(\mathcal{R})}(2\partial_{(\alpha} f'(\mathcal{R}) 
\delta^\gamma_{\beta)}-g^{\gamma\sigma}g_{\alpha\beta}\partial_\sigma f'(\mathcal{R}))\,;
\ee
we can find two independent expressions by contracting respectively $\alpha$ and $\beta$ indices or $\alpha$ and $\gamma$ in the last equation; the 
resulting conditions being
\be
Q_\gamma-2\widetilde{Q}_\gamma=-\frac{2}{f'(\mathcal{R})}\partial_\gamma f'(\mathcal{R})\,,
\ee
\be
-Q_\beta=\frac{4}{f'(\mathcal{R})}\partial_\beta f'(\mathcal{R})\,.
\ee
It is now clear that both the constraint $Q_\mu=0$ or $\widetilde{Q}_\mu=0$ lead to the same result 
$Q_\mu=\widetilde{Q}_\mu=\partial_\mu f'(\mathcal{R})=0$, that makes the theory obviously inconsistent since it forces the function $f(\mathcal{R})$
to be at most linear.

A viable alternative \cite{Vitagliano:2010sr,Sotiriou:2006qn} is the theory obtained constraining the trace of the torsion tensor $S_{\mu\rho}{}^\rho$ 
through the term $C^\mu S_{\mu}\equiv C^\mu S_{\mu\rho}{}^\rho$
\bea
\mathcal{S}=\frac{1}{16\pi L_P^2}\int d^4x \det( f(\mathcal{R})+C^\mu S_\mu)+\mathcal{S}_M(g_{\mu\nu}, \Gamma^\gamma{}_{\alpha\beta}, \psi)\,,
\eea
whose field equations are written as
\bea\label{fe1}
\hspace{-2.4cm}f'(\mathcal{R}) \mathcal{R}_{(\mu\nu)}-\frac{1}{2}f(\mathcal{R})g_{\mu\nu}=(8\pi L_P^2) T_{\mu\nu}\,,\nonumber
\eea
\bea\label{fe2}
\hspace{-2.4cm}-\s{\Gamma}{\nabla}_{\lambda}(\sqrt{-g}f'(\mathcal{R})g^{\mu\nu})+\s{\Gamma}{\nabla}_{\sigma}\l(\sqrt{-g}f'(\mathcal{R})g^{\sigma\mu}\r)\delta^{\nu}_{\;\;\lambda}+\nn\\ 
\hspace{-2.18cm}+2\sqrt{-g}f'(\mathcal{R})(g^{\mu\nu}S_{\lambda\sigma}^{\;\;\;\;\sigma}-g^{\mu\rho}\delta^{\nu}_{\;\;\lambda}S_{\rho\sigma}^{\;\;\;\;\sigma}
+g^{\mu\sigma}S_{\sigma\lambda}^{\;\;\;\;\nu})=(8\pi L_P^2)\sqrt{-g} \l(\Delta^{\mu\nu}{}_{\lambda}
-\frac{2}{3}\Delta^{\sigma[\nu}{}_{\sigma}\delta^{\mu]}_{\;\;\lambda}\r)\!,\nn
\eea
\be
\label{fe3}
\hspace{-2.4cm}S_{\alpha\mu}^{\;\;\;\;\alpha}=0\,.
\ee
Note that this choice is fully consistent: if we require vacuum solutions, the second of equations (\ref{fe3}) simply reduces to the two conditions 
$S_{\mu\nu\rho}=0$ and $\s{\Gamma}{\nabla}_{\lambda}(\sqrt{-g}f'(\mathcal{R})g^{\mu\nu})=0$, that are the usual field equations found in Palatini-$f(\mathcal{R})$ 
theories of gravity. On the other side, this version of $f(\mathcal{R})$-metric-affine theories has the further feature to avoid propagation of torsion 
waves in vacuum. In fact (modulo the dependence of the matter Lagrangian on the covariant derivative, that must be at most linear), taking the 
antisymmetric part of second equation in (\ref{fe3}) with respect to $\mu$ and $\nu$, and adding suitable permutations of the obtained expression, 
we can show that torsion tensor is
\be
S_{\mu\nu}{}^\lambda=\frac{8\pi L_P^2}{f'(\mathcal{R})}g^{\rho\lambda}(\Delta_{[\rho\mu]\nu}+\Delta_{[\nu\rho]\mu}-\Delta_{[\mu\nu]\rho})\,,
\ee
namely, the torsion tensor is algebraically defined by the antisymmetric part of the hypermomentum tensor, $\Delta^{[\mu\nu]}{}_\lambda$.

\section{Conclusions}

Metric-affine theories of gravity are among the most valuable generalizations of Einstein gravity. The most appealing characteristic of these models is
to make straightforward the possibility of a coupling between the geometry of the spacetime (through the independent connection) and the internal degrees
of freedom of matter fields ({\em viz}. intrinsic particle spin, dilation current and shear). While the dynamics of the torsion tensor has been extensively 
studied in literature starting from the simplest case of the Einstein-Cartan theory till higher order theories, the role of nonmetricity has been 
usually shelved. 

In this paper we have explored the dynamics of metric-affine theory of gravity including both torsion and nonmetricity. We have shown that, in the most 
general theory obtained at the second order through dimensional analysis, the independent connection is algebraically expressed in terms of matter 
fields, metric and their derivatives. Torsion and nonmetricity gain dynamics only when higher order Lagrangian densities are taken into account, as can 
be proved including a Ricci-squared correction to the Einstein-Cartan action. We have also reanalyzed the rather peculiar case of $f(\mathcal{R})$ theories of 
gravity, where the projective invariance of the gravitational Lagrangian forces the introduction of a further constraint on the connection. We showed 
that the attempt of gauge fixing the degrees of freedom introduced by the projective transformation via nonmetricity leads always to inevitable 
inconsistencies. 

A possible way to bypass the problem of the projective invariance is to constrain some of the degrees of freedom of torsion by adding an appropriate 
Lagrange multiplier. The field equations obtained from the re-arranged theory result to be fully consistent. 
This theory represents the metric-affine generalization of modified theories of gravity, whose phenomenology surely deserves further investigation. In 
particular it would be very interesting to understand what is the behaviour of the extra degrees of freedom residing in the connection when they are 
excited, and how matter interactions are modified when energies are well below the dynamical regime of connection. These topics will be pursued in future 
publications.

\section*{Acknowledgements}
The Author would like to thank Roberto Percacci and Yuri Obukhov for the suggestion to explore the role of nonmetricity in the dynamics of MAG.
Also, a special acknowledgement to Andrea Nerozzi for his help with the xAct package \cite{xact}, used to check some of the
calculations in the paper. Finally, the Author is grateful to Antonino Flachi and Stefano Liberati for a critical reading of the manuscript.
VV is supported by FCT - Portugal through the grant SFRH/BPD/77678/2011.
% 
% \section*{Appendix}
% 
% The field equation obtained varying with respect to the metric the action (\ref{general}) is the following
% \bea
% \mathcal{R}_{\alpha\beta}-\frac{1}{2}g_{\alpha\beta}\mathcal{R}+\nn\\
% c_1 (S_{\alpha\gamma}{}^{\gamma} S_{\beta\delta}{}^{\delta} -  \frac{1}{2} g_{\alpha\beta} S_{\gamma\delta}{}^{\delta} \
% S^{\gamma}{}_{\epsilon}{}^{\epsilon})+c_2(S_{\alpha\gamma}{}^{\delta} S_{\beta\delta}{}^{\gamma} -  \frac{1}{2} g_{\alpha\beta} S_{\gamma\delta}{}^{\epsilon} \
% S^{\gamma}{}_{\epsilon}{}^{\delta})+\nn\\
% +c_3(2 S_{\alpha\gamma}{}^{\delta} S_{\beta}{}^{\gamma}{}_{\delta} -  S_{\gamma\delta\alpha} S^{\gamma\delta}{}_{\beta} -  \frac{1}{2} g_{\alpha\beta} S_{\gamma\delta}{}^{\epsilon} S^{\gamma\delta}{}_{\epsilon})
% \eea
% 

%%%%%%%%%%%%%%%%%%%%%%%%%%%%%%%%%%%%%%%%%%%%%%%%%%%%%%%%%%%%%%%%%%%%%%%%%%%%%%%%%%%%%%%%%%%%%%%%%%%%%%%%%%%%%%%%%%%%%

\end{document}